\documentstyle[12pt]{article}

\begin{document}  

\title{\Large {\bf Quantum motion equation and Poincar{\' e} 
               translation invariance of noncommutative field theory} }

\author{{{\large Zheng Ze Ma}} \thanks{Electronic address: 
           z.z.ma@seu.edu.cn} 
  \\  \\ {\normalsize {\sl Department of Physics, Southeast University, 
         Nanjing, 210096, P. R. China } }}

\date{}

\maketitle

\begin{abstract}

  We study the Moyal commutators and their expectation values between 
vacuum states and non-vacuum states for noncommutative scalar field 
theory. For noncommutative $\varphi^{\star4}$ scalar field theory, we 
derive its energy-momentum tensor from translation transformation and 
Lagrange field equation. We generalize the Heisenberg and quantum 
motion equations to the form of Moyal star-products for noncommutative 
$\varphi^{\star4}$ scalar field theory for the case $\theta^{0i}=0$ of 
spacetime noncommutativity. Then we demonstrate the Poincar{\' e} 
translation invariance for noncommutative $\varphi^{\star4}$ scalar 
field theory for the case $\theta^{0i}=0$ of spacetime noncommutativity. 

\end{abstract}

~~ PACS numbers: 11.10.Nx, 11.30.Cp

\section{Introduction}

\indent 

  The researches of gravitation theory and superstring theories in recent 
years reveal that spacetime coordinates may be noncommutative under 
certain microscopic scales such as the Planck scale [1-4]. This has 
confirmed the concept of spacetime quantization 
proposed by Snyder many years ago [5]. For quantum field theories, they 
are set up in ordinary commutative spacetime. Because spacetime may be 
noncommutative, people need to study quantum field theories on 
noncommutative spacetime. A lot of research works have been carried out on 
noncommutative field theories in recent years. However for noncommutative 
field theories, there are still many problems need to be studied further.

  Poincar{\' e} translation invariance and Lorentz rotation invariance 
are the basic spacetime symmetries for quantum field theories on ordinary 
commutative spacetime. For quantum field theories on noncommutative 
spacetime, we need to study whether these properties are still satisfied. 
According to superstring theories, people usually take $\theta^{\mu\nu}$ 
to be constant $c$-number matrix that not changed under reference system 
transformations. This will destroy the Lorentz rotation invariance 
generally for noncommutative field theory, except for certain special 
forms of $\theta^{\mu\nu}$, the invariance under a subgroup 
$SO(1,1)\times SO(2)$ of the usual Lorentz group is reserved [6]. 
In Refs. [7,8] the authors put forward the twisted Poincar{\' e} algebra 
explanation for the relativistic invariance of noncommutative field 
theories. However the quantum realization of such a proposition is not 
clear yet. In Ref. [9] the authors take $\theta^{\mu\nu}$ to be a 
$c$-number second-order antisymmetric tensor and then demonstrated the  
Poincar{\' e} translation invariance and Lorentz rotation invariance 
for noncommutative field theories from their classical field equation 
approach.

  In this paper, we study the quantum motion equations and Poincar{\' e} 
translation invariance of noncommutative field theory. For simplicity, 
we only study the noncommutative scalar field theory. In Sec. II, we study 
the Moyal commutators and their expectation values between vacuum states 
and non-vacuum states for noncommutative scalar field theory. In Sec. III, 
we derive the energy-momentum tensor for noncommutative $\varphi^{\star4}$ 
scalar field theory from translation transformation and Lagrange field 
equation. In Sec. IV, we generalize the Heisenberg and quantum motion 
equations to the form of Moyal star-products for noncommutative 
$\varphi^{\star4}$ scalar field theory and then demonstrate its 
Poincar{\' e} translation invariance from the Heisenberg and quantum 
motion equation approach. However we need the condition $\theta^{0i}=0$
for the spacetime noncommutativity. For the case $\theta^{0i}\neq0$ of 
spacetime noncommutativity, it seems that Poincar{\' e} translation 
invariance for noncommutative scalar field theory cannot be set up from 
Heisenberg and quantum motion equation approach. In Sec. V, we discuss 
some of the problems.

\section{ Commutators of the Moyal star-products and their 
          expectation values }

\indent

  According to the researches of gravitation and superstring 
theories [1,2], we know that spacetime may have noncommutative structures 
under the Planck scale. The noncommutativity of spacetime can be realized 
through introducing the commutation relation
$$
  [x^{\mu},x^{\nu}]=i\theta^{\mu\nu} 
  \eqno{(2.1)}  $$
for the spacetime coordinates, where $\theta^{\mu\nu}$ is a real 
antisymmetric matrix that parameterizes the noncommutativity of the 
spacetime. $\theta^{\mu\nu}$ has the dimension of square of length. 
For quantum field theories on noncommutative spacetime, they can be 
realized through introducing the Moyal star-product, i.e., all of the 
products between field functions or field operators that depend on 
spacetime coordinates are replaced by the Moyal star-products. The 
Moyal star-product of two fields $f(x)$ and $g(x)$ is defined to be 
\begin{eqnarray*}
  f(x)\star g(x) & = & e^{\frac{i}{2}\theta^{\mu\nu}\frac{\partial}{\partial 
                 \alpha^\mu}\frac{\partial}{\partial\beta^{\nu}}}f(x+\alpha)
                 g(x+\beta)\vert_{\alpha=\beta=0} \\
                 & = & f(x)g(x)+\sum\limits^{\infty}_{n=1}
                 \left(\frac{i}{2}\right)^{n}
                 \frac{1}{n!}\theta^{\mu_{1}\nu_{1}}
                 \cdots\theta^{\mu_{n}\nu_{n}}
                 \partial_{\mu_{1}}\cdots\partial_{\mu_{n}}f(x) 
                 \partial_{\nu_{1}}\cdots\partial_{\nu_{n}}g(x) ~.
\end{eqnarray*}
$$ 
  \eqno{(2.2)}  $$

  In noncommutative spacetime, the Moyal star-product can be regarded as 
the fundamental product operation. Therefore it is necessary to study 
the commutators of Moyal star-products for quantum fields on noncommutative 
spacetime. The Lagrangian for the noncommutative $\varphi^{\star4}$ scalar 
field theory is given by 
$$
  {\cal L}=\frac{1}{2}\partial^{\mu}\varphi\star\partial_{\mu}\varphi-
           \frac{1}{2}m^{2}\varphi\star\varphi-\frac{1}{4!}\lambda
             \varphi\star\varphi\star\varphi\star\varphi ~.
  \eqno{(2.3)}  $$
The Fourier expansion of the free scalar field is given by 
$$
  \varphi({\bf x},t)=\int\frac{d^{3}k}{\sqrt{(2\pi)^{3}2\omega_{k}}}
       [a(k)e^{i{\bf k}\cdot{\bf x}-i\omega t}+
        a^{\dagger}(k)e^{-i{\bf k}\cdot{\bf x}+i\omega t}] ~.
  \eqno{(2.4)}  $$
Here we adopt the usual Lorentz invariant form for the Fourier expansion,  
hence it is just the same as the case of the commutative spacetime, except 
that in Eq. (2.4) we take the spacetime coordinates to be noncommutative 
which satisfy the commutation relation (2.1). It is reasonable to take the 
commutation relations for the creation and annihilation operators to be the 
same as the commutative spacetime: 
$$
  [a(k),a^{\dagger}(k^{\prime})]=\delta^{3}({\bf k}-{\bf k}^{\prime}) ~,  $$
$$ 
  [a(k),a(k)]=0 ~,      $$
$$
  [a^{\dagger}(k),a^{\dagger}(k)]=0 ~.              
  \eqno{(2.5)}  $$

  The Moyal star-product of two field functions of Eq. (2.2) is defined 
at the same spacetime point. We can generalize Eq. (2.2) to define the 
Moyal star-product of two field functions on different spacetime 
points [4]:
\begin{eqnarray*}
  f(x_{1})\star g(x_{2}) & = & e^{\frac{i}{2}
          \theta^{\mu\nu}\frac{\partial}{\partial 
          \alpha^\mu}\frac{\partial}{\partial\beta^{\nu}}}f(x_{1}+\alpha)
                 g(x_{2}+\beta)\vert_{\alpha=\beta=0} \\
                 & = & f(x_{1})g(x_{2})+\sum\limits^{\infty}_{n=1}
                 \left(\frac{i}{2}\right)^{n}
                 \frac{1}{n!}\theta^{\mu_{1}\nu_{1}}
                 \cdots\theta^{\mu_{n}\nu_{n}}
                 \partial_{\mu_{1}}\cdots\partial_{\mu_{n}}f(x_{1}) 
                 \partial_{\nu_{1}}\cdots\partial_{\nu_{n}}g(x_{2}) ~,
\end{eqnarray*}
$$ 
  \eqno{(2.6)}  $$
which means that the commutation relation of spacetime coordinates (2.1) 
is generalized to arbitrary two different spacetime points: 
$$
  [x_{1}^{\mu},x_{2}^{\nu}]=i\theta^{\mu\nu} ~. 
  \eqno{(2.7)}  $$
We define the commutator of two scalar fields of Moyal star-product to 
be 
$$
  [\varphi(x),\varphi(y)]_{\star}=\varphi(x)\star\varphi(y)-
              \varphi(y)\star\varphi(x) ~.
  \eqno{(2.8)}  $$
We can call Eq. (2.8) the Moyal commutator for convenience. 
From the Fourier expansion of Eq. (2.4) for the scalar field, we can 
calculate the Moyal commutator for two scalar fields. It is given by 
\begin{eqnarray*}
  [\varphi(x),\varphi(y)]_{\star}
           & = & \int\frac{d^{3}kd^{3}k^{\prime}}
                {(2\pi)^{3}\sqrt{2\omega_{k}2\omega_{k}^{\prime}}} 
                \left[a(k)e^{-ikx}+a^{\dagger}(k)e^{ikx},  
                 a(k^{\prime})e^{-ik^{\prime}y}+a^{\dagger}
                 (k^{\prime})e^{ik^{\prime}y}\right]_\star      \\ 
           & = & \int\frac{d^{3}kd^{3}k^{\prime}}
         {(2\pi)^{3}\sqrt{2\omega_{k}2\omega_{k}^{\prime}}}\Bigg\{
      \left[a(k)e^{-ikx},a(k^{\prime})e^{-ik^{\prime}y}\right]_\star+
      \left[a(k)e^{-ikx},a^{\dagger}(k^{\prime})e^{ik^{\prime}y}
               \right]_\star               \\
           & ~ & ~~~~~~~~~ +\left[a^{\dagger}(k)e^{ikx},a(k^{\prime})
                      e^{-ik^{\prime}y}\right]_\star+
                 \left[a^{\dagger}(k)e^{ikx},a^{\dagger}
                 (k^{\prime})e^{ik^{\prime}y}\right]_\star\Bigg\} ~.
\end{eqnarray*}      
$$  \eqno{(2.9)}  $$
For the reason that there exist two kinds of noncommutative objects, 
i.e., field operators and spacetime coordinates, the spacetime coordinates 
now is noncommutative, we cannot apply the commutation relations for the 
creation and annihilation operators of Eq. (2.5) directly to obtain a 
$c$-number result for the Moyal commutator.

  In order to obtain a $c$-number result for the Moyal commutator, we can 
calculate its vacuum state expectation value. We have 
\begin{eqnarray*}
   & ~ & \langle0\vert[\varphi(x),\varphi(y)]
            _{\star}\vert0\rangle              \\
           & = & \langle0\vert\int\frac{d^{3}kd^{3}k^{\prime}}
            {(2\pi)^{3}\sqrt{2\omega_{k}2\omega_{k}^{\prime}}}   
        \left(a(k)a^{\dagger}(k^{\prime})e^{-ikx}\star e^{ik^{\prime}y}   
         -a(k^{\prime})a^{\dagger}(k)e^{-ik^{\prime}y}\star e^{ikx}
          \right)\vert0\rangle   \\
           & = & \int\frac{d^{3}k}{(2\pi)^{3}2\omega_{k}}   
                 \left[e^{-ikx}\star e^{iky}-
                  e^{-iky}\star e^{ikx}\right]       \\ 
           & = & \int\frac{d^{3}k}{(2\pi)^{3}2\omega_{k}}   
                 \left[\exp(\frac{i}{2}k\times k)e^{-ik(x-y)}
                   -\exp(\frac{i}{2}k\times k)e^{ik(x-y)}\right] ~,  
\end{eqnarray*}        
$$  \eqno{(2.10)}  $$
where we have applied Eq. (2.6) and we note 
$$
  k\times p=k_{\mu}\theta^{\mu\nu}p_{\nu} ~.
  \eqno{(2.11)}  $$
Because $\theta^{\mu\nu}$ is antisymmetric, $k\times k=0$, we obtain
\begin{eqnarray*}
  & ~ & \langle0\vert[\varphi(x),\varphi(y)]_{\star}\vert0\rangle
        =\int\frac{d^{3}k}{(2\pi)^{3}2\omega_{k}}   
                 \left[e^{-ik(x-y)}-e^{ik(x-y)}\right]            \\
  & = & -\frac{i}{(2\pi)^{3}}\int\frac
                    {d^{3}k}{\omega_{k}}e^{i{\bf k}\cdot({\bf x}-
       {\bf y})}\sin\omega_{k}(x_{0}-y_{0}) = i\Delta(x-y) ~. 
\end{eqnarray*}      
$$ 
  \eqno{(2.12)}  $$
We can see that the result of Eq. (2.12) is just equal to the commutator 
of two scalar fields in ordinary commutative spacetime: 
$$
  [\varphi(x),\varphi(y)]=-\frac{i}{(2\pi)^{3}}\int\frac
           {d^{3}k}{\omega_{k}}e^{i{\bf k}\cdot({\bf x}-{\bf y})}
           \sin\omega_{k}(x_{0}-y_{0}) = i\Delta(x-y) ~.      
  \eqno{(2.13)}  $$ 
It is obvious to see that this equality relies on the antisymmetry of 
$\theta^{\mu\nu}$.

  The vacuum state expectation value of the equal-time Moyal commutator 
of two scalar fields can be obtained from Eq. (2.12):
$$
  \langle0\vert[\varphi({\bf x},t),\varphi({\bf y},t)]_{\star}
          \vert0\rangle=\Delta({\bf x}-{\bf y},0)=0 ~.       
  \eqno{(2.14)}  $$ 
The time derivative of the function $\Delta$ at the origin of the 
coordinates is given by 
$$
  \frac{\partial\Delta(x-y)}{\partial x_{0}}\Bigg\vert_{x_{0}=y_{0}}
           =-\delta^{3}({\bf x}-{\bf y}) ~.
  \eqno{(2.15)}  $$ 
The conjugate momentum for the scalar field is defined to be 
$$
  \pi=\frac{\partial{\cal L}}{\partial\stackrel{\cdot}{\varphi}}
         =\stackrel{\cdot}{\varphi} ~.
  \eqno{(2.16)}  $$ 
From Eqs. (2.12) and (2.16) we can obtain the following vacuum state 
expectation value for the equal-time Moyal commutators: 
$$
  \langle0\vert[\pi({\bf x},t),\pi({\bf y},t)]_{\star}
        \vert0\rangle=0 ~,                
  \eqno{(2.17)}  $$ 
$$
  \langle0\vert[\pi({\bf x},t),\varphi({\bf y},t)]_{\star}\vert0\rangle=
                  -i\delta^{3}({\bf x}-{\bf y}) ~.
  \eqno{(2.18)}  $$

  We can also calculate the expectation values between non-vacuum states 
for the Moyal commutators. Let $\vert\Psi\rangle$ represent a normalized 
non-vacuum physical state which is in the occupation eigenstate:
$$
  \vert\Psi\rangle=\vert N_{k_{1}}N_{k_{2}}\cdots N_{k_{i}}\cdots,0
              \rangle ~,
  \eqno{(2.19)}  $$ 
where $N_{k_{i}}$ represents the occupation number of the momentum $k_{i}$. 
We can suppose that the occupation numbers are nonzero only on some 
separate momentums $k_{i}$. For all other momentums, the occupation 
numbers are zero. We use $0$ to represent that the occupation numbers are 
zero on all the other momentums in Eq. (2.19). The state vector 
$\vert\Psi\rangle$ has the following properties: 
$$  
  \langle N_{k_{1}}N_{k_{2}}\cdots N_{k_{i}}\cdots\vert
          N_{k_{1}}N_{k_{2}}\cdots N_{k_{i}}\cdots\rangle =1 ~,     
  \eqno{(2.20)}  $$ 
$$
  \sum\limits_{N_{k_{1}}N_{k_{2}}\cdots}
    \vert N_{k_{1}}N_{k_{2}}\cdots N_{k_{i}}\cdots\rangle
    \langle N_{k_{1}}N_{k_{2}}\cdots N_{k_{i}}\cdots\vert={\rm I} ~,
  \eqno{(2.21)}  $$   
$$
  a(k_{i})\vert N_{k_{1}}N_{k_{2}}\cdots N_{k_{i}}\cdots\rangle=
          \sqrt{N_{k_{i}}}\vert N_{k_{1}}N_{k_{2}}\cdots 
                (N_{k_{i}}-1)\cdots\rangle ~,           
  \eqno{(2.22)}  $$ 
$$
  a^{\dagger}(k_{i})\vert N_{k_{1}}N_{k_{2}}\cdots N_{k_{i}}
          \cdots\rangle=\sqrt{N_{k_{i}}+1}\vert N_{k_{1}}N_{k_{2}}
          \cdots(N_{k_{i}}+1)\cdots\rangle ~.
  \eqno{(2.23)}  $$ 
Equation (2.21) is the completeness expression for the state vector 
$\vert\Psi\rangle$. Therefore Eq. (2.19) can represent an arbitrary 
scalar field quantum sysytem. 
Then from Eq. (2.9) we can obtain the expectation value between any 
non-vacuum physical states for the Moyal commutator (2.8) to be 
\begin{eqnarray*}
  & ~ & \langle\Psi\vert[\varphi(x),\varphi(y)]_{\star}
               \vert\Psi\rangle              \\
  & = & \langle\Psi\vert\int\frac{d^{3}kd^{3}k^{\prime}}
         {(2\pi)^{3}\sqrt{2\omega_{k}2\omega_{k}^{\prime}}}\left\{
          \left[a(k)e^{-ikx},a^{\dagger}(k^{\prime})e^{ik^{\prime}y}
               \right]_\star              
        +\left[a^{\dagger}(k)e^{ikx},a(k^{\prime})
         e^{-ik^{\prime}y}\right]_\star\right\}\vert\Psi\rangle   \\
  & = & \int\frac{d^{3}k}{(2\pi)^{3}2\omega_{k}}   
                 \left[e^{-ikx}\star e^{iky}-
                 e^{-iky}\star e^{ikx} \right]  
            =i\Delta(x-y) ~,
\end{eqnarray*}  
$$  \eqno{(2.24)}  $$
which is just equal to the vacuum state expectation value of Eq. (2.12). 
For the equal-time commutators, we can also obtain 
$$
  \langle\Psi\vert[\varphi({\bf x},t),\varphi({\bf y},t)]_{\star}
          \vert\Psi\rangle=\Delta({\bf x}-{\bf y},0)=0 ~,       
  \eqno{(2.25)}  $$ 
$$
  \langle\Psi\vert[\pi({\bf x},t),\pi({\bf y},t)]_{\star}
        \vert\Psi\rangle=0 ~,                
  \eqno{(2.26)}  $$ 
$$
  \langle\Psi\vert[\pi({\bf x},t),\varphi({\bf y},t)]_{\star}
         \vert\Psi\rangle=-i\delta^{3}({\bf x}-{\bf y}) ~.
  \eqno{(2.27)}  $$ 
The properties of the commutation relations for the creation and 
annihilation operators of Eq. (2.5) are still reflected in the above 
evaluations for the non-vacuum state expectation values for the Moyal 
commutators.

  Because $\Delta(x-y)$ is a Lorentz invariant singular function, 
it has the property 
$$
  \Delta(x-y)=0 ~~~~~~~~ \mbox{for} ~~~~~~~~ (x-y)^{2}<0 ~.
  \eqno{(2.28)}  $$
This means  
$$
  \langle0\vert[\varphi(x),\varphi(y)]_{\star}\vert0\rangle=0
           ~~~~~~~~ \mbox{for} ~~~~~~~~ (x-y)^{2}<0 ~, 
  \eqno{(2.29)}  $$
and similarly  
$$
  \langle\Psi\vert[\varphi(x),\varphi(y)]_{\star}
             \vert\Psi\rangle=0
           ~~~~~~~~ \mbox{for} ~~~~~~~~ (x-y)^{2}<0 ~. 
  \eqno{(2.30)}  $$
For quantum field theories on ordinary spacetime, the commutators 
satisfy the microscopic causality. For scalar field we have  
$$
  [\varphi(x),\varphi(y)]=0
           ~~~~~~~~ \mbox{for} ~~~~~~~~ (x-y)^{2}<0 ~. 
  \eqno{(2.31)}  $$
This means that any two fields as physical observables commute with 
each other when they are separated by a spacelike interval. Thus any 
two fields as physical observables at different spacetime points can 
be measured precisely and independently of each other only if they 
cannot be connected by a light signal or any other physical information. 
For scalar field theory on noncommutative spacetime, although we cannot 
conclude that $[\varphi(x),\varphi(y)]_{\star}=0$ for $(x-y)^{2}<0$ from 
Eqs. (2.29) and (2.30), Eqs. (2.29) and (2.30) still represent the 
satisfying of microscopic causality for scalar field theory on 
noncommutative spacetime. This is because any physical measurement is 
taken under certain physical state. What the observer measures are all 
expectation values in fact.

However in the above we have only analyzed the microcausality 
property for the linear operator $\varphi(x)$. For the quadratic 
operators of free scalar field on noncommutative spacetime such 
as $\varphi(x)\star\varphi(x)$, their microcausality properties 
need to be studied further. Some of their results have been obtained 
in Refs. [10,11]. In addition, the microcausality problem 
discussed here is only restricted to free fields. Because for 
noncommutative field theories, there exist the UV/IR mixing 
problems [12,13]. The infrared singularities that come from 
non-planar diagrams may need one to invoke certain nonlocal terms in 
the renormalization of noncommutative field theories. These nonlocal 
terms may destroy the microcausality for quantum field 
theories on noncommutative spacetime.

\section{ The energy-momentum tensor of noncommutative 
                     $\varphi^{\star4}$ scalar field theory }

\indent

  In ordinary commutative field theories, Poincar{\' e} translation 
invariance results the existence of locally conserved energy-momentum 
tensors. In this section we will derive the energy-momentum tensor for 
noncommutative $\varphi^{\star4}$ scalar field theory from translation 
transformation and Lagrange field equation. The Lagrangian for 
noncommutative $\varphi^{\star4}$ scalar field theory is given by 
Eq. (2.3). To consider an infinitesimal displacements of spacetime 
coordinates 
$$
  x^{\prime\mu}=x^{\mu}+\epsilon^{\mu} ~, 
  \eqno{(3.1)}  $$ 
we can see in Eq. (2.1), because $\theta^{\mu\nu}$ does not depend on 
the coordinates, the spacetime noncommutative relations (2.1) and (2.7) 
are translation invariant. The Lagrangian of Eq. (2.3) does not depend 
on the coordinates explicitly, hence it is translation invariant. For 
the Lagrangian (2.3) in the form of Moyal star-products, it only contains 
the fields and their first order derivative, therefore we have 
$$
  {\cal L}={\cal L}(\varphi,\partial\varphi/\partial x_{\mu}) ~.  
  \eqno{(3.2)}  $$
The field equation can still be obtained from the variation principle 
formally. It is given by 
$$
  \frac{\partial}{\partial x_{\mu}}\frac{\partial{\cal L}}
       {\partial(\partial\varphi/\partial x_{\mu})}-
           \frac{\partial{\cal L}}{\partial\varphi}=0 ~.
  \eqno{(3.3)}  $$
To substitute Eq. (2.3) into Eq. (3.3), we obtain 
$$
  (\Box+m^{2})\varphi+\frac{1}{3!}\lambda\varphi\star\varphi
              \star\varphi=0 ~.
  \eqno{(3.4)}  $$
Under the infinitesimal displacements of Eq. (3.1), the Lagrangian will 
have a little displacement: 
$$ 
  \delta{\cal L}={\cal L}(x^{\prime})-{\cal L}(x)
           =\epsilon_{\mu}\frac{\partial{\cal L}}{\partial x_{\mu}} ~.
  \eqno{(3.5)}  $$
On the other hand, from Eq. (3.2) we have
$$
  \delta{\cal L}=\frac{\partial{\cal L}}{\partial\varphi}
           \star\delta\varphi+\frac{\partial{\cal L}}
          {\partial(\partial\varphi/\partial x_{\mu})}\star
          \delta\left(\frac{\partial\varphi}{\partial x_{\mu}}\right) ~,
  \eqno{(3.6)}  $$
where 
$$
  \delta\varphi=\varphi(x+\epsilon)-\varphi(x)=
        \epsilon_{\mu}\frac{\partial\varphi(x)}{\partial x_{\mu}} ~.
  \eqno{(3.7)}  $$
From Eqs. (3.5), (3.6), and (3.3), we obtain 
$$
  \epsilon_{\mu}\frac{\partial{\cal L}}{\partial x_{\mu}}=
    \frac{\partial}{\partial x_{\mu}}\left(\frac{\partial{\cal L}}
    {\partial(\partial\varphi/\partial x_{\mu})}\star
    \epsilon_{\nu}\frac{\partial\varphi}{\partial x_{\nu}}\right) ~.
  \eqno{(3.8)}  $$
Because Eq. (3.8) is satisfied for arbitrary $\epsilon_{\mu}$, we obtain 
$$
  \frac{\partial{\cal T}_{\mu\nu}}{\partial x_{\mu}}=0 ~,
  \eqno{(3.9)}  $$
where ${\cal T}_{\mu\nu}$ is the energy-momentum tensor defined to be 
$$
  {\cal T}_{\mu\nu}=-g_{\mu\nu}{\cal L}+\frac{\partial{\cal L}}
       {\partial(\partial\varphi/\partial x_{\mu})}\star
       \frac{\partial\varphi}{\partial x_{\nu}} ~.
  \eqno{(3.10)}  $$
To notice that the Moyal star-products are not invariant generally 
under the commutation of the orders of two functions, we need to write 
Eq. (3.10) in a symmetrized form for the Moyal star-products: 
$$
  {\cal T}_{\mu\nu}=-g_{\mu\nu}{\cal L}+\frac{1}{2}\left(
       \frac{\partial{\cal L}}{\partial(\partial\varphi/\partial x_{\mu})}
       \star\frac{\partial\varphi}{\partial x_{\nu}}
       +\frac{\partial\varphi}{\partial x_{\nu}}\star
        \frac{\partial{\cal L}}{\partial(\partial\varphi/
        \partial x_{\mu})}\right) ~.
  \eqno{(3.11)}  $$
In fact this can be resulted through symmetrizing the orders of the Moyal 
star-products between two functions properly in the above derivation. 
To substitute Eq. (2.3) into Eq. (3.11) we obtain
$$
  {\cal T}_{\mu\nu}=\frac{1}{2}(\partial_{\mu}\varphi\star
          \partial_{\nu}\varphi+\partial_{\nu}\varphi\star
          \partial_{\mu}\varphi)-g_{\mu\nu}{\cal L} ~.
  \eqno{(3.12)}  $$ 
In Refs. [14,15], Eq. (3.12) has been obtained from the translation 
invariance of the action. Here we derive it from a different method.

  However the divergence of the energy-momentum tensor obtained above is 
not zero. In fact the divergence of the energy-momentum tensor of 
Eq. (3.12) is given by [14,15] 
$$
  \partial^{\mu}{\cal T}_{\mu\nu}=\frac{\lambda}{4!}\left[\left[\varphi,
            \partial_{\nu}\varphi\right]_{\star},
            \varphi^{\star 2}\right]_{\star} ~.
  \eqno{(3.13)}  $$
Therefore the energy-momentum tensor of Eq. (3.12) is not locally 
conserved. From the integral property of the Moyal star-product, 
${\cal T}_{\mu\nu}$ is conserved to the integral over the whole 
spacetime:
$$
  \partial^{\mu}\int d^{4}x{\cal T}_{\mu\nu}=\int d^{4}x 
           \partial^{\mu}{\cal T}_{\mu\nu}=0 ~.               
  \eqno{(3.14)}  $$
Equation (3.14) can also be resulted from the Gauss theorem. 
We can analyze why we cannot obtain a locally conserved energy-momentum 
tensor from the above derivation. The reason is that: the Lagrangian (2.3) 
contains higher order derivative terms when it is 
expanded according to the Moyal star-product (2.2). If we expand the 
Lagrangian completely, it will contain higher order derivative terms up 
till infinite order. We may write it in the form 
$$
  {\cal L}={\cal L}(\varphi,\partial\varphi/\partial x_{\mu},
            \partial^{2}\varphi/\partial x_{\mu}\partial x_{\nu},\cdots ) ~, 
  \eqno{(3.15)}  $$
where the star-products are canceled with the ordinary products left. 
In such a case, the Lagrange field equation is not Eq. (3.3) in fact, 
and the variation of the Lagrangian under infinitesimal displacements of 
Eq. (3.1) is not Eq. (3.6). Therefore the above derivation of the 
energy-momentum tensor for the Lagrangian (2.3) given in the form of Moyal 
star-products is only formal. Thus we cannot obtain a locally conserved 
energy-momentum tensor from such a method. Perhaps one can obtain a 
locally conserved energy-momentum tensor from the completely expanded 
Lagrangian of the Moyal star-products. However because it will contain 
infinite terms, it is not realizable actually. In Ref. [16] the action 
for noncommutative gauge field theory is expanded to the first order 
of the parameter $\theta^{\mu\nu}$ and the energy-momentum tensor to 
the first order of $\theta^{\mu\nu}$ has been derived. However 
the energy-momentum tensor obtained there is not locally conserved, 
excepting to define a modified energy-momentum tensor. We consider that 
the reason is related with the fact the total Lagrangian of a 
noncommutative field theory contains infinite terms when it is expanded 
according to the Moyal star-product. In Ref. [17] the action for 
noncommutative gauge field theory is expanded to the second order 
of the parameter $\theta^{\mu\nu}$.

  Although the energy-momentum tensor of Eq. (3.12) is not locally 
conserved, we can still define such a four-momentum $P_{\mu}$ formally:
$$
  P_{\mu}=\int d^{3}x{\cal T}_{0\mu}~.
  \eqno{(3.16)}  $$   
From Eq. (3.12) we have 
$$
  H=\int d^{3}x{\cal T}_{00}=\int d^{3}x (\pi\star\pi-{\cal L})
         =\int d^{3}x(\pi\star\stackrel{\cdot}{\varphi}-{\cal L}) ~,
  \eqno{(3.17)}  $$
$$
  P_{i}=\int d^{3}x{\cal T}_{0i}=\int d^{3}x\frac{1}{2}
        (\pi\star\partial_{i}\varphi+\partial_{i}\varphi\star\pi) ~.
  \eqno{(3.18)}  $$
We use the Bjorken metric throughout this paper [18]. Hence the 
three-dimensional momentum is ${\bf P}=(-P_{1},-P_{2},-P_{3})$. 
From Eq. (3.17), we can see that $H$ is equal to the definition of 
the total energy of noncommutative scalar field theory in the form 
of Moyal star-products. Because 
$$  
  \int d^{3}x\partial^{\mu}{\cal T}_{\mu\nu}=\partial^{0}\int d^{3}x
            {\cal T}_{0\nu}+\int d^{3}x\partial^{i}{\cal T}_{i\nu}=
             \partial^{0}\int d^{3}x{\cal T}_{0\nu} ~,
  \eqno{(3.19)}  $$
we obtain 
$$
  \partial^{0}P_{\mu}=\int d^{3}x\frac{\lambda}{4!}\left[\left[\varphi,
            \partial_{\mu}\varphi\right]_{\star},
            \varphi^{\star 2}\right]_{\star} ~.
  \eqno{(3.20)}  $$  
Because the right hand side of Eq. (3.20) is not zero generally, we 
cannot define a conserved four-momentum $P_{\mu}$ for noncommutative 
$\varphi^{\star4}$ scalar field theory.

  For the case $\theta^{0i}=0$ of spacetime noncommutativity, a totally 
conserved energy-momentum four-vector can be defined for noncommutative 
$\varphi^{\star4}$ scalar field theory [14,15]. This is because the 
integral of Moyal star-products on three-dimensional space satisfies the 
cyclic property  
$$
  \int d^{3}x(\varphi_{1}\star\varphi_{2}\star\cdots\star\varphi_{n})(x) 
          =\int d^{3}x(\varphi_{n}\star\varphi_{1}\star\cdots
           \star\varphi_{n-1})(x) 
  \eqno{(3.21)}  $$
when $\theta^{0i}=0$. Therefore we have
$$
  \partial^{0}P_{\mu}=\int d^{3}x\frac{\lambda}{4!}\left[\left[\varphi,
            \partial_{\mu}\varphi\right]_{\star},
            \varphi^{\star 2}\right]_{\star}=0 
  \eqno{(3.22)}  $$  
for the case $\theta^{0i}=0$. In Sec. IV, we will demonstrate the 
Poincar{\' e} translation invariance for noncommutative 
$\varphi^{\star4}$ scalar field theory from the Heisenberg and quantum 
motion equation approach. However this will need the condition that the 
energy-momentum $P_{\mu}$ are conserved quantities. Because for the 
case $\theta^{0i}=0$ of spacetime noncommutativity, the energy-momentum 
$P_{\mu}$ are conserved quantities, this make us possible to establish 
the Poincar{\' e} translation invariance for noncommutative 
$\varphi^{\star4}$ scalar field theory from Heisenberg and quantum 
motion equation approach for the case $\theta^{0i}=0$ of spacetime 
noncommutativity. From the above analysis we can see that for 
noncommutative field theories, although the Lagrangians and actions 
may be translation invariant, it may not necessarily result the 
existence of locally conserved energy-momentum tensors.

\section{ Quantum motion equation and Poincar{\' e} translation 
              invariance for noncommutative scalar field theory }

\subsection{ Heisenberg equation on noncommutative spacetime }

\indent

  The fundamental evolution equation for quantum fields in ordinary 
commutative spacetime is Heisenberg equation. The Heisenberg equations 
for the scalar field and its conjugate momentum are 
$$
  i[H,\varphi(x)]=\frac{\partial\varphi(x)}{\partial t} ~,
  \eqno{(4.1)}  $$
$$
  i[H,\pi(x)]=\frac{\partial\pi(x)}{\partial t} ~.
  \eqno{(4.2)}  $$
The solutions of Eqs. (4.1) and (4.2) can be given by the following 
formal integrals:
$$
  \varphi({\bf x},t)=e^{iHt}\varphi({\bf x},0)e^{-iHt} ~,
  \eqno{(4.3)}  $$
$$
  \pi({\bf x},t)=e^{iHt}\pi({\bf x},0)e^{-iHt} ~.
  \eqno{(4.4)}  $$
For quantum field theories on noncommutative spacetime, such as 
noncommutative scalar field theory, its Hamiltonian can be defined by 
$$
  H=\int d^{3}x{\cal H}
   =\int d^{3}x(\pi\star\stackrel{\cdot}{\varphi}-{\cal L}) ~, 
  \eqno{(4.5)}  $$
where ${\cal L}$ is the Lagrangian on noncommutative spacetime. For  
noncommutative $\varphi^{\star4}$ scalar field theory, its Lagrangian is 
given by Eq. (2.3). Furthermore to consider that the Moyal star-products 
are not invariant generally under the commutation of the orders of two 
functions, we need to write Eq. (4.5) in a symmetrized form for fields 
and their conjugate momentums such as 
$$
  H=\int d^{3}x{\cal H}
   =\int d^{3}x\left(\frac{1}{2}\pi\star\stackrel{\cdot}{\varphi}+
      \frac{1}{2}\stackrel{\cdot}{\varphi}\star\pi-{\cal L}\right) ~.
  \eqno{(4.6)}  $$
Because in Eqs. (4.1) and (4.2), the commutators are the usual 
commutators in commutative spacetime, for quantum field theories 
on noncommutative spacetime, in order to make Eqs. (4.1) and (4.2) 
as the fundamental evolution equations for fields and their conjugate 
momentums, we need to expand the Hamiltonian according to the Moyal 
star-product of Eq. (2.2) in principle. Supposing that we have 
expanded the Hamiltonian of Eq. (4.6) according to the Moyal 
star-product (2.2), then we can still regard Eqs. (4.1) and (4.2) as 
the fundamental evolution equations for fields and their conjugate 
momentums on noncommutative spacetime.

  We can find that the formal solutions of Eqs. (4.1) and (4.2) can also 
be given by the following integrals: 
$$
  \varphi({\bf x},t)=e^{iHt}\star\varphi({\bf x},0)\star e^{-iHt} ~,
  \eqno{(4.7)}  $$
$$
  \pi({\bf x},t)=e^{iHt}\star\pi({\bf x},0)\star e^{-iHt} ~,
  \eqno{(4.8)}  $$
i.e., in Eqs. (4.3) and (4.4), we replace the ordinary products by the 
Moyal star-products. This can be verified through substituting Eqs. 
(4.7) and (4.8) into Eqs. (4.1) and (4.2). On the other hand, we can see 
that Heisenberg equations (4.1) and (4.2) can be rewritten in the form 
$$
  i[H,\varphi(x)]_\star=\frac{\partial\varphi(x)}{\partial t} ~,
  \eqno{(4.9)}  $$
$$
  i[H,\pi(x)]_\star=\frac{\partial\pi(x)}{\partial t} ~.
  \eqno{(4.10)}  $$
This is because we can consider that the integral of Hamiltonian density 
over the whole space in Eq. (4.6) has been carried out first, thus $H$ 
does not rely on the space variables. On the other hand, because $H$ 
is the total energy of the system, we can think that it does not rely 
on the time variable. This makes 
$$
  H\cdot\varphi(x)=H\star\varphi(x)  ~~~~~~~~ \mbox{and} ~~~~~~~~
  \varphi(x)\cdot H=\varphi(x)\star H ~,                      $$
$$
  H\cdot\pi(x)=H\star\pi(x)  ~~~~~~~~ \mbox{and} ~~~~~~~~
  \pi(x)\cdot H=\pi(x)\star H                                
  \eqno{(4.11)}  $$
according to the definition of the Moyal star-product (2.2). Thus we have 
$$
  [H,\varphi(x)]=[H,\varphi(x)]_\star   ~~~~~~~~ \mbox{and} ~~~~~~~~ 
  [H,\pi(x)]=[H,\pi(x)]_\star ~.                              
  \eqno{(4.12)}  $$
This makes the establishment of Eqs. (4.9) and (4.10). In fact we can see 
that Eqs. (4.7) and (4.8) are also the formal integrals of Eqs. (4.9) and 
(4.10), under the condition that $H$ does not rely on the spacetime 
variables. This means that Heisenberg equations (4.1) and (4.2) can be 
generalized to the form of noncommutative spacetime directly. However we 
have seen in Sec. III that for noncommutative field theories, the total 
energy of the system may not be a conserved quantity generally. This 
means that the Hamiltonian $H$ may depend on the time variable. For the 
case that $H$ depends on the time variable, the satisfying of Eqs. (4.11) 
and (4.12) need the condition $\theta^{0i}=0$ for the spacetime 
noncommutativity. Under such a condition, the formal integrals of 
Eqs. (4.9) and (4.10) can still be given by Eqs. (4.7) and (4.8). 
Therefore for noncommutative field theories of general case, the 
generalization of Heisenberg equations (4.1) and (4.2) to the form of 
Eqs. (4.9) and (4.10) need the condition $\theta^{0i}=0$ of the 
spacetime noncommutativity.

\subsection{ Heisenberg relations on noncommutative spacetime }

\indent

  Quantum field theories in ordinary spacetime satisfy the 
Poincar{\' e} translation invariance. To consider an infinitesimal 
displacement of the spacetime coordinates    
$$
  x^{\prime\mu}=x^{\mu}+\epsilon^{\mu} ~, 
\eqno{(4.13)}$$
whose generation operator is given by 
$$
  U(\epsilon)=\exp(i\epsilon_{\mu}P^{\mu})
              \approx1+i\epsilon_{\mu}P^{\mu} ~,
\eqno{(4.14)}$$
where $P^{\mu}$ are the generators of the displacement transformation. 
Under the displacement transformation (4.13), the scalar field satisfies  
$$
  U(\epsilon)\varphi(x)U(\epsilon)^{-1}=\varphi(x+\epsilon) ~.
\eqno{(4.15)}$$
From Eq. (4.15) we obtain 
$$
  i[P^{\mu},\varphi(x)]=\frac{\partial\varphi(x)}{\partial x_{\mu}} ~,
\eqno{(4.16)}$$ 
which is the Heisenberg relations that should be satisfied for the 
Poincar{\' e} translation invariance of quantum field theory. In 
Eq. (4.16), $P^{\mu}$ are constructed from the energy-momentum tensor. 
Because $P^{\mu}$ are the generators of the displacement transformation, 
they should be conserved quantities of motion. For quantum field 
theories on ordinary spacetime, Eq. (4.16) is verified through the 
equal-time commutation relations and Lagrange field equations [18].

  For quantum field theories on noncommutative spacetime, if they 
satisfy the Poincar{\' e} translation invariance, it need to be 
guaranteed same by the Heisenberg relations (4.16). However the 
four-momentum $P^{\mu}$ should be conserved quantities of motion. 
But we have seen in Sec. III that it is difficult to define a locally 
conserved energy-momentum tensor and hence a conserved four-momentum 
for noncommutative field theories generally. If $P^{\mu}$ are not the 
conserved quantities of motion, the satisfying of Hesenberg relations 
has no meaning. However we have seen in Sec. III that for noncommutative 
$\varphi^{\star4}$ scalar field theory, for the case $\theta^{0i}=0$ of 
spacetime noncommutativity, the four-momentum $P^{\mu}$ defined through 
Eqs. (3.16) to (3.18) are conserved quantities of motion. Therefore for 
such a case, we are possible to verify the Poincar{\' e} translation 
invariance from the Hesenberg relations (4.16).

  As in Eqs. (4.1) and (4.2), the commutators in Eq. (4.16) are 
commutators of ordinary products. Thus in order to verify the 
Poincar{\' e} translation invariance from Eq. (4.16), we need to expand 
the four-momentum $P^{\mu}$ according to the Moyal star-product. However 
there will be infinite expansion terms and we cannot carry out to obtain 
them completely in fact. Therefore we hope to keep the products between 
the field operators to be the Moyal star-products in $P^{\mu}$ and to 
apply the Moyal commutators to verify Eq. (4.16) for quantum field 
theories on noncommutative spacetime.

  In Eq. (4.16), if we consider that the integrals of the four-momentum 
densities over the whole space have been carried out first, 
then $P^{\mu}$ do not rely on the space variables. At the same time, if 
$P^{\mu}$ are conserved quantities of motion, then $P^{\mu}$ do not rely 
on the time variable. Therefore the products between $P^{\mu}$ and 
$\varphi(x)$ have the property 
$$
  P^{\mu}\cdot\varphi(x)=P^{\mu}\star\varphi(x) ~, ~~~~~~~~
  \varphi(x)\cdot P^{\mu}=\varphi(x)\star P^{\mu}             
  \eqno{(4.17)}  $$
according to the definition of the Moyal star-product (2.2) if 
we consider that the space integral operations have been performed 
first. However for Eq. (4.17), we can see that we need the condition 
$\theta^{0i}=0$ if the four-momentum $P^{\mu}$ are not the conserved 
quantities which depend on time. But we know that for noncommutative 
field theories, the four-momentum $P^{\mu}$ that obtained and defined 
from the method of Sec. III are not conserved quantities of motion 
generally. Therefore we need the condition $\theta^{0i}=0$ of  
spacetime noncommutativity for the relation (4.17) for noncommutative 
field theories generally. Under such a condition, equation (4.16) can be 
rewritten in the form 
$$
  i[P^{\mu},\varphi(x)]_{\star}=\frac{\partial\varphi(x)}
            {\partial x_{\mu}} 
\eqno{(4.18)}$$ 
equivalently according to the relation (4.17). This means that it is 
possible for us to verify the Poincar{\' e} translation invariance for 
noncommutative field theories from the Moyal commutators, i.e., to keep  
the products between field operators to be the Moyal star-products in 
$P^{\mu}$ and not to expand them according to Eq. (2.2). And we know 
that under the condition $\theta^{0i}=0$, the four-momentum $P^{\mu}$ 
defined through Eqs. (3.16) to (3.18) are conserved quantities of motion 
for noncommutative $\varphi^{\star4}$ scalar field theory.

\subsection{ Moyal commutators for coupling fields }

\indent

  For quantum field theories on ordinary spacetime, Heisenberg relations 
(4.16) are verified through the equal-time commutation relations. For 
scalar field theory these equal-time commutation relations are  
$$
  [\varphi({\bf x},t),\varphi({\bf y},t)]=0 ~,       
  \eqno{(4.19)}  $$ 
$$
  [\pi({\bf x},t),\pi({\bf y},t)]=0 ~,                
  \eqno{(4.20)}  $$ 
$$
  [\pi({\bf x},t),\varphi({\bf y},t)]=
                  -i\delta^{3}({\bf x}-{\bf y}) ~.
  \eqno{(4.21)}  $$ 
These equal-time commutation relations are first established for free 
fields, then through Heisenberg equations they are proved to be true 
also for coupling fields [18]. For the commutators of Moyal 
star-products, as seen in Sec. II, they are not $c$-number results. 
In order to obtain the $c$-number results for the Moyal commutators, 
we need to evaluate their vacuum state or non-vacuum state expectation 
values. For scalar field, we have obtained in Sec. II the expectation 
values for the Moyal commutators. For the sake of convenience, we 
write down them here again:
$$
  \langle\Psi\vert[\varphi({\bf x},t),\varphi({\bf y},t)]_{\star}
          \vert\Psi\rangle=0 ~,       
  \eqno{(4.22)}  $$ 
$$
  \langle\Psi\vert[\pi({\bf x},t),\pi({\bf y},t)]_{\star}
        \vert\Psi\rangle=0 ~,                
  \eqno{(4.23)}  $$ 
$$
  \langle\Psi\vert[\pi({\bf x},t),\varphi({\bf y},t)]_{\star}
         \vert\Psi\rangle=-i\delta^{3}({\bf x}-{\bf y}) ~.
  \eqno{(4.24)}  $$ 
Here we use $\vert\Psi\rangle$ to represent the vacuum state and 
non-vacuum state together. As shown in Sec. II, these relations are 
established for free fields at this time.

  We hope to verify that these relations are also satisfied for 
coupling fields. Like that for quantum field theories on ordinary 
commutative spacetime, we can suppose that at time $t=0$, the 
interactions have not participated, the interactions participate 
adiabaticly at time $t>0$. Then at time $t=0$ the field operators 
$\varphi({\bf x},0)$ and $\pi({\bf x},0)$ can be expanded in the 
form of free fields: 
$$
  \varphi({\bf x},0)=\int\frac{d^{3}k}{\sqrt{(2\pi)^{3}2\omega_{k}}}
       [a(k)e^{i{\bf k}\cdot{\bf x}}+
        a^{\dagger}(k)e^{-i{\bf k}\cdot{\bf x}}] ~, 
  \eqno{(4.25)}  $$
$$
  \pi({\bf x},0)=\stackrel{\cdot}{\varphi}({\bf x},0)
        =\int\frac{d^{3}k}{\sqrt{(2\pi)^{3}2\omega_{k}}}
         (-i\omega)[a(k)e^{i{\bf k}\cdot{\bf x}}-
          a^{\dagger}(k)e^{-i{\bf k}\cdot{\bf x}}] ~.  
  \eqno{(4.26)}  $$
To suppose that the expansion coefficients $a^{\dagger}(k)$ and $a(k)$ 
satisfy the same commutation relations (2.5) of free fields, this will 
make the coupling fields satisfy the relations (4.22) to (4.24) at time 
$t=0$:
$$
  \langle\Psi\vert[\varphi({\bf x},0),\varphi({\bf y},0)]_{\star}
          \vert\Psi\rangle=0 ~,       
  \eqno{(4.27)}  $$ 
$$
  \langle\Psi\vert[\pi({\bf x},0),\pi({\bf y},0)]_{\star}
        \vert\Psi\rangle=0 ~,                
  \eqno{(4.28)}  $$ 
$$
  \langle\Psi\vert[\pi({\bf x},0),\varphi({\bf y},0)]_{\star}
         \vert\Psi\rangle=-i\delta^{3}({\bf x}-{\bf y}) ~.
  \eqno{(4.29)}  $$
These relations can be resulted through repeating the procedure of 
Sec. II. Or equivalently we can regard that the fields are in the free 
propagating state from time $t=-\infty$ to $t=0$. Then the coupling 
fields satisfy the relations (4.27) to (4.29) at time $t=0$ as a 
corollary of the results of Sec. II.

  Now we need to verify that the relations (4.27) to (4.29) are also 
satisfied for coupling fields at an arbitrary time. To take Eq. (4.29) 
for example, we product it with $e^{iEt}$ from the left hand side and 
product it with $e^{-iEt}$ from the right hand side:
$$
  e^{iEt}\langle\Psi\vert[\pi({\bf x},0),\varphi({\bf y},0)]_{\star}
     \vert\Psi\rangle e^{-iEt}=e^{iEt}
     [-i\delta^{3}({\bf x}-{\bf y})]e^{-iEt}
    =-i\delta^{3}({\bf x}-{\bf y}) ~,
  \eqno{(4.30)}  $$
where $E$ is the total energy of the state $\vert\Psi\rangle$.
As postulated in Sec. II, the state vector $\vert\Psi\rangle$ is in 
the occupation eigenstate, we can suppose that it is also in the 
energy eigenstate at the same time. Hence we have 
$$ 
  \langle\Psi\vert e^{iHt}=\langle\Psi\vert e^{iEt}        
   ~~~~~~~~ \mbox{and} ~~~~~~~~
  e^{-iHt}\vert\Psi\rangle=e^{-iEt}\vert\Psi\rangle ~.
  \eqno{(4.31)}  $$
To apply Eq. (4.31) we have 
\begin{eqnarray*}
  & ~ & e^{iEt}\langle\Psi\vert[\pi({\bf x},0),\varphi({\bf y},0)]
         _{\star}\vert\Psi\rangle e^{-iEt}        \\
  & =& \langle\Psi\vert e^{iHt}
     [\pi({\bf x},0),\varphi({\bf y},0)]_{\star}e^{-iHt}
     \vert\Psi\rangle       \\
  & = & \langle\Psi\vert[e^{iHt}\pi({\bf x},0)e^{-iHt},
        e^{iHt}\varphi({\bf y},0)e^{-iHt}]_{\star}\vert\Psi\rangle ~,  
\end{eqnarray*}
$$  \eqno{(4.32)}  $$
where in the third line, we have taken $H$ as being integrated 
therefore it does not depend on the space coordinates. Then from 
Eqs. (4.3) and (4.4), we obtain 
$$
  \langle\Psi\vert[\pi({\bf x},t),\varphi({\bf y},t)]_{\star}
       \vert\Psi\rangle=-i\delta^{3}({\bf x}-{\bf y})  
  \eqno{(4.33)}  $$ 
for the coupling fields at an arbitrary time. For Eqs. (4.27) and 
(4.28), we can also obtain  
$$
  \langle\Psi\vert[\varphi({\bf x},t),\varphi({\bf y},t)]_{\star}
          \vert\Psi\rangle=0       
  \eqno{(4.34)}  $$ 
and 
$$
  \langle\Psi\vert[\pi({\bf x},t),\pi({\bf y},t)]_{\star}
        \vert\Psi\rangle=0                
  \eqno{(4.35)}  $$ 
for the coupling fields at an arbitrary time. To mention here that in 
the above derivation, the state vector $\vert\Psi\rangle$ is an arbitrary 
state vector for the scalar field system including the vacuum state, 
while not necessarily to be the state of the interacting and evolutional 
quantum field system at the time $t$. Of course the relations (4.33) to 
(4.35) are also satisfied for the vacuum state for coupling fields at an 
arbitrary time: 
$$
  \langle0\vert[\pi({\bf x},t),\varphi({\bf y},t)]_{\star}
       \vert0\rangle=-i\delta^{3}({\bf x}-{\bf y}) ~,
  \eqno{(4.36)}  $$ 
$$
  \langle0\vert[\varphi({\bf x},t),\varphi({\bf y},t)]_{\star}
          \vert0\rangle=0 ~,       
  \eqno{(4.37)}  $$ 
$$
  \langle0\vert[\pi({\bf x},t),\pi({\bf y},t)]_{\star}
        \vert0\rangle=0 ~.               
  \eqno{(4.38)}  $$ 
However in the following demonstration, what we need are relations 
(4.36) to (4.38), while not (4.33) to (4.35).

  Another relation for the equal-time Moyal commutators that we need to 
use in the verifying of the Poincar{\' e} translation invariance for 
noncommutative scalar field theory is 
$$
  \langle0\vert[\partial_{i}\varphi({\bf x},t),\varphi({\bf y},t)]
        _{\star}\vert0\rangle=0 ~.
  \eqno{(4.39)}  $$ 
The corresponding relation of Eq. (4.39) in ordinary commutative 
spacetime is 
$$
  [\partial_{i}\varphi({\bf x},t),\varphi({\bf y},t)]=0 ~.
  \eqno{(4.40)}  $$ 
For the case of the free field, Eq. (4.39) can be calculated like that 
for Eq. (2.10). We need not to write down the calculation completely. 
We only write down its last expression here:
$$
  \langle0\vert[\partial_{i}\varphi({\bf x},t),\varphi({\bf y},t)]
         _{\star}\vert0\rangle
        =\int\frac{d^{3}k}{(2\pi)^{3}2\omega_{k}}(ik_{i}) 
         \left[e^{i{\bf k}\cdot({\bf x}-{\bf y})}
          +e^{-i{\bf k}\cdot({\bf x}-{\bf y})}\right] ~,           
  \eqno{(4.41)}  $$
where we note ${\bf k}=(k_{1},k_{2},k_{3})$. In Eq. (4.41) because 
the integrand now is an odd function, this makes the whole integral 
to be zero. Then through the argument of this subsection, we can 
obtain that Eq. (4.41) is also held for coupling fields.

\subsection{ Poincar{\' e} translation invariance }

\indent

  Now we can verify the Poincar{\' e} translation invariance for 
noncommutative $\varphi^{\star4}$ scalar field theory from the 
Heisenberg relations (4.18) on noncommutative spacetime. From Eqs. 
(3.17) and (3.18), we have 
$$
  P^{0}=H=\int d^{3}x{\cal H}(\pi,\varphi) ~,             $$
\begin{eqnarray*}
  {\cal H}(\pi,\varphi) & = & 
       \frac{1}{2}\Big[\pi({\bf x},t)\star\pi({\bf x},t)+
  \partial_{i}\varphi({\bf x},t)\star\partial_{i}\varphi({\bf x},t) \\ 
  & ~ & ~~~ + m^{2}\varphi({\bf x},t)\star\varphi({\bf x},t)+
        \frac{2}{4!}\lambda\varphi({\bf x},t)\star\varphi({\bf x},t)
        \star\varphi({\bf x},t)\star\varphi({\bf x},t)\Big] ~,
\end{eqnarray*}
$$  \eqno{(4.42)}  $$
and 
$$
  {\bf P}=(P^{1},P^{2},P^{3}) ~,                 $$
$$
  P^{i}=-\int d^{3}x\frac{1}{2}[\pi({\bf x},t)\star
         \partial_{i}\varphi({\bf x},t)+\partial_{i}
         \varphi({\bf x},t)\star\pi({\bf x},t)] ~.
  \eqno{(4.43)}  $$

  We first verify the space components of Eq. (4.18). For the three 
space components of Eq. (4.18), they are in the equal positions. Thus 
we need to check only one of them. We take the $x^{1}$-component for 
example. We have 
$$
  i[P^{1},\varphi({\bf x}^{\prime},t)]_{\star}=
        -\frac{1}{2}\int d^{3}x[\pi({\bf x},t)\star
         \partial_{1}\varphi({\bf x},t)+\partial_{1}
         \varphi({\bf x},t)\star\pi({\bf x},t),
         \varphi({\bf x}^{\prime},t)]_{\star} ~.
  \eqno{(4.44)}  $$
To use the formula 
$$
  [a\star b,c]_{\star}=a\star[b,c]_{\star}
               +[a,c]_{\star}\star b ~, 
  \eqno{(4.45)}  $$
we have 
$$
  [\pi({\bf x},t)\star\partial_{1}\varphi({\bf x},t),
   \varphi({\bf x}^{\prime},t)]_{\star}=\pi({\bf x},t)\star
   [\partial_{1}\varphi({\bf x},t),\varphi({\bf x}^{\prime},t)]
   _{\star}+[\pi({\bf x},t),\varphi({\bf x}^{\prime},t)]_{\star}
   \star\partial_{1}\varphi({\bf x},t) ~,               $$
$$
  [\partial_{1}\varphi({\bf x},t)\star\pi({\bf x},t),
         \varphi({\bf x}^{\prime},t)]_{\star}=\partial_{1}
     \varphi({\bf x},t)\star[\pi({\bf x},t),\varphi
     ({\bf x}^{\prime},t)]_{\star}+[\partial_{1}\varphi({\bf x},t),
     \varphi({\bf x}^{\prime},t)]_{\star}\star\pi({\bf x},t) ~.
  \eqno{(4.46)}  $$    
For the reason that the Moyal commutators are not $c$-number functions, 
we can evaluate the vacuum expectation values for Eqs. (4.44) and (4.46). 
We have 
\begin{eqnarray*}
  & ~ & \langle0\vert[\pi({\bf x},t)\star\partial_{1}\varphi({\bf x},t),
   \varphi({\bf x}^{\prime},t)]_{\star}\vert0\rangle     \\
  & = & \langle0\vert\pi({\bf x},t)\star
   [\partial_{1}\varphi({\bf x},t),\varphi({\bf x}^{\prime},t)]
   _{\star}\vert0\rangle+\langle0\vert
    [\pi({\bf x},t),\varphi({\bf x}^{\prime},t)]_{\star}
   \star\partial_{1}\varphi({\bf x},t)\vert0\rangle ~,              
\end{eqnarray*}
\begin{eqnarray*}
  & ~ & \langle0\vert[\partial_{1}\varphi({\bf x},t)\star\pi({\bf x},t),
         \varphi({\bf x}^{\prime},t)]_{\star}\vert0\rangle    \\
  & = & \langle0\vert\partial_{1}\varphi({\bf x},t)\star
         [\pi({\bf x},t),\varphi({\bf x}^{\prime},t)]
         _{\star}\vert0\rangle+\langle0\vert
        [\partial_{1}\varphi({\bf x},t),\varphi({\bf x}^{\prime},t)]
        _{\star}\star\pi({\bf x},t)\vert0\rangle ~.
\end{eqnarray*}
$$  \eqno{(4.47)}  $$    
In order to use the previous result for the Moyal commutators, we insert 
the unit operator $\vert0\rangle\langle0\vert=\mbox{I}$ inside Eq. (4.47). 
Thus we obtain 
\footnote{Here, we can suppose the state vector $\vert0\rangle$ is 
in the momentum representation, therefore it does not rely on the 
coordinates, we have 
$\vert0\rangle\langle0\vert=\vert0\rangle\star\langle0\vert$. To be 
more reasonable, we may insert the complete set of the state vectors 
such as that of Eq. (2.21). We rewrite Eq. (2.21) as 
$$  \sum\limits_{N_{k_{1}}N_{k_{2}}\cdots}
    \vert N_{k_{1}}N_{k_{2}}\cdots N_{k_{i}}\cdots\rangle
    \langle N_{k_{1}}N_{k_{2}}\cdots N_{k_{i}}\cdots\vert =  
 \vert0\rangle\langle0\vert + 
 \sum\limits_{M=1}^{\infty}\vert M\rangle\langle M\vert        
  = \vert0\rangle\star\langle0\vert + 
 \sum\limits_{M=1}^{\infty}\vert M\rangle\star\langle M\vert ~.  $$ 
Here we use $\vert1\rangle$ to represent the state vectors that the 
occupation numbers on each momentum are no more than one, however at 
least one of the occupation number is one. We use $\vert2\rangle$ to 
represent the state vectors that the occupation numbers on each 
momentum are no more than two, however at least one of the occupation 
number is two. Similarly for $M\geq3$. We have supposed that the state 
vectors $\vert M\rangle$ are in the momentum representation, 
therefore they do not rely on the coordinates, and this makes 
$\vert M\rangle\langle M\vert=\vert M\rangle\star\langle M\vert$. 
In the above expression, we 
have omitted a summation index for each $\vert M\rangle$. We can obtain 
that in Eq. (4.48), the contributions that come from the state vectors 
$\vert1\rangle$, $\vert2\rangle$, ... are all zero. 
For example, we can obtain 
$\langle1\vert[\partial_{1}\varphi({\bf x},t),
\varphi({\bf x}^{\prime},t)]_{\star}\vert0\rangle=0$, 
$\langle0\vert\pi({\bf x},t)\vert2\rangle=0$, 
$\langle3\vert[\partial_{1}\varphi({\bf x},t),
\varphi({\bf x}^{\prime},t)]_{\star}\vert0\rangle=0$, 
$\langle0\vert\pi({\bf x},t)\vert3\rangle=0$, etc.. 
For free fields it is so. For coupling fields, it is also as such. 
We omit to write down the detailed analysis here. Therefore we can 
only insert the expression 
$\vert0\rangle\langle0\vert=\mbox{I}$ equivalently.}   
\begin{eqnarray*}
  & ~ & \langle0\vert[\pi({\bf x},t)\star\partial_{1}\varphi({\bf x},t),
   \varphi({\bf x}^{\prime},t)]_{\star}\vert0\rangle     \\
  & = & \langle0\vert\pi({\bf x},t)\vert0\rangle\star\langle0\vert
   [\partial_{1}\varphi({\bf x},t),\varphi({\bf x}^{\prime},t)]
   _{\star}\vert0\rangle+\langle0\vert
    [\pi({\bf x},t),\varphi({\bf x}^{\prime},t)]_{\star}\vert0\rangle
   \star\langle0\vert\partial_{1}\varphi({\bf x},t)\vert0\rangle ~,              
\end{eqnarray*}
\begin{eqnarray*}
  & ~ & \langle0\vert[\partial_{1}\varphi({\bf x},t)\star\pi({\bf x},t),
         \varphi({\bf x}^{\prime},t)]_{\star}\vert0\rangle    \\
  & = & \langle0\vert\partial_{1}\varphi({\bf x},t)
        \vert0\rangle\star\langle0\vert
         [\pi({\bf x},t),\varphi({\bf x}^{\prime},t)]
         _{\star}\vert0\rangle+\langle0\vert
        [\partial_{1}\varphi({\bf x},t),\varphi({\bf x}^{\prime},t)]
        _{\star}\vert0\rangle\star\langle0\vert
         \pi({\bf x},t)\vert0\rangle ~.
\end{eqnarray*}
$$  \eqno{(4.48)}  $$ 
To apply Eqs. (4.36) and (4.39), we obtain 
$$
  \langle0\vert[\pi({\bf x},t)\star\partial_{1}\varphi({\bf x},t),
   \varphi({\bf x}^{\prime},t)]_{\star}\vert0\rangle=
  -i\delta^{3}({\bf x}-{\bf x}^{\prime})\star\langle0\vert
    \partial_{1}\varphi({\bf x},t)\vert0\rangle ~,              $$
$$
  \langle0\vert[\partial_{1}\varphi({\bf x},t)\star\pi({\bf x},t),
         \varphi({\bf x}^{\prime},t)]_{\star}\vert0\rangle=-i
  \langle0\vert\partial_{1}\varphi({\bf x},t)
        \vert0\rangle\star\delta^{3}({\bf x}-{\bf x}^{\prime}) ~.
  \eqno{(4.49)}  $$ 
To substitute Eq. (4.49) into Eq. (4.44) we have 
\begin{eqnarray*}
  & ~ & \langle0\vert i[P^{1},\varphi({\bf x}^{\prime},t)]_{\star}
     \vert0\rangle       \\  
& = & -\frac{1}{2}\int d^{3}x
     [\delta^{3}({\bf x}-{\bf x}^{\prime})\star\langle0\vert
      \partial_{1}\varphi({\bf x},t)\vert0\rangle+
      \langle0\vert\partial_{1}\varphi({\bf x},t)
        \vert0\rangle\star\delta^{3}({\bf x}-{\bf x}^{\prime})]   \\ 
       & = & \frac{1}{2}\langle0\vert\int d^{3}x
     [\delta^{3}({\bf x}-{\bf x}^{\prime})\star
      \partial^{1}\varphi({\bf x},t)+\partial^{1}\varphi({\bf x},t)
        \star\delta^{3}({\bf x}-{\bf x}^{\prime})]\vert0\rangle ~.
\end{eqnarray*}    
$$  \eqno{(4.50)}  $$
To compare the two sides of Eq. (4.50), we can move the vacuum states 
away. Thus we obtain 
$$
  i[P^{1},\varphi({\bf x}^{\prime},t)]_{\star}=
      \frac{1}{2}\int d^{3}x[\delta^{3}({\bf x}-{\bf x}^{\prime})
      \star\partial^{1}\varphi({\bf x},t)+\partial^{1}\varphi({\bf x},t)
        \star\delta^{3}({\bf x}-{\bf x}^{\prime})] ~.
  \eqno{(4.51)}  $$
Under the condition $\theta^{0i}=0$ of the spacetime noncommutativity, 
we have the formula
$$
  \int d^{3}x\delta^{3}({\bf x}-{\bf x}^{\prime})\star f({\bf x},t)=
  \int d^{3}x f({\bf x},t)\star\delta^{3}({\bf x}-{\bf x}^{\prime})=
   f({\bf x}^{\prime},t) ~. 
  \eqno{(4.52)}  $$ 
A derivation for Eq. (4.52) is given in the Appendix. To apply Eq. (4.52) 
in Eq. (4.51), we obtain 
$$
  i[P^{1},\varphi({\bf x}^{\prime},t)]_{\star}=
       \partial^{1}\varphi({\bf x}^{\prime},t) ~.
  \eqno{(4.53)}  $$ 
Or equivalently, we write Eq. (4.53) as  
$$
  i[P^{1},\varphi({\bf x},t)]_{\star}=\frac
         {\partial\varphi({\bf x},t)}{\partial x_{\mu}} ~.
  \eqno{(4.54)}  $$ 
Thus we have finished the proof for the $x$-component of the Heisenberg 
relations (4.18). For the $y$- and $z$-component of the Heisenberg 
relations (4.18), because they are the same as the $x$-component, we 
need not to repeat it.

  Now we need to check the time component of the Heisenberg relations 
(4.18). The method is the same as for the $x$-component. To use Eq. 
(4.45) and to insert the unit operator 
$\vert0\rangle\star\langle0\vert=\mbox{I}$,
\footnote{The same reason at that of Eq. (4.48), we need to insert 
the complete set of the state vectors. However the contribution that 
come from non-vacuum states are all zero in Eq. (4.55). Hence we can 
only insert the expression $\vert0\rangle\langle0\vert=\mbox{I}$ 
equivalently. Because the state vector $\vert0\rangle$ is in the 
momentum representation, it does not rely on the coordinates, we have 
$\vert0\rangle\langle0\vert=\vert0\rangle\star\langle0\vert$.}
then to use Eqs. (4.37) and (4.39), we have 
$$
  \langle0\vert[\partial_{i}\phi({\bf x},t)\star\partial_{i}
   \phi({\bf x},t),\varphi({\bf x}^{\prime},t)]
    _\star\vert0\rangle=0 ~,          $$
$$
  \langle0\vert[\varphi({\bf x},t)\star\varphi({\bf x},t), 
   \varphi({\bf x}^{\prime},t)]_\star\vert0\rangle=0 ~,          $$
\begin{eqnarray*}
  & ~ & \langle0\vert[\varphi({\bf x},t)\star\varphi({\bf x},t)
        \star\varphi({\bf x},t)\star\varphi({\bf x},t),
    \varphi({\bf x}^{\prime},t)]_\star\vert0\rangle    \\
  & = & \langle0\vert \varphi({\bf x},t)\star\varphi({\bf x},t)       
     \star[\varphi({\bf x},t)\star\varphi({\bf x},t),
     \varphi({\bf x}^{\prime},t)]_\star\vert0\rangle    \\
  & ~ & + \langle0\vert[\varphi({\bf x},t)\star\varphi({\bf x},t),
     \varphi({\bf x}^{\prime},t)]_\star\star\varphi({\bf x},t)
     \star\varphi({\bf x},t)\vert0\rangle =0 ~.
\end{eqnarray*}
$$  \eqno{(4.55)}  $$
To use Eq. (4.45) we obtain 
\begin{eqnarray*}
  & ~ & \langle0\vert i[H,\varphi({\bf x}^{\prime},t)]_{\star}
     \vert0\rangle         \\
   & = & \frac{i}{2}\langle0\vert\int d^{3}x           
     [\pi({\bf x},t)\star\pi({\bf x},t),\
      \varphi({\bf x}^{\prime},t)]_{\star}\vert0\rangle    \\
   & = & \frac{i}{2}\int d^{3}x \left(\langle0\vert
       [\pi({\bf x},t),\varphi({\bf x}^{\prime},t)]_{\star}
       \star\pi({\bf x},t)\vert0\rangle +
       \langle0\vert\pi({\bf x},t)\star[\pi({\bf x},t),
       \varphi({\bf x}^{\prime},t)]_{\star}\vert0\rangle\right) ~.   
\end{eqnarray*}
$$  \eqno{(4.56)}  $$
To insert the unit operator $\vert0\rangle\star\langle0\vert=\mbox{I}$, 
the same reason as that of Eqs. (4.48) and (4.55), then to use 
Eqs. (4.36) and (4.52), we obtain 
$$
  \langle0\vert i[H,\varphi({\bf x}^{\prime},t)]_{\star}\vert0\rangle  
  =\langle0\vert \pi({\bf x}^{\prime},t)\vert0\rangle  
  =\langle0\vert \stackrel{\cdot}{\varphi}
    ({\bf x}^{\prime},t)\vert0\rangle ~.    
  \eqno{(4.57)}  $$
To move away the vacuum states in both sides of Eq. (4.57) and to 
replace ${\bf x}^{\prime}$ by ${\bf x}$, we obtain 
$$
  i[H,\varphi({\bf x},t)]_{\star}=
   \frac{\partial\varphi({\bf x},t)}{\partial t} ~.    
  \eqno{(4.58)}  $$
Thus we have proved the time component of the Heisenberg relations (4.18).

  All together we have proved that for noncommutative $\varphi^{\star4}$ 
scalar field theory, under the condition $\theta^{0i}=0$ of the spacetime 
noncommutativity, $\varphi(x)$ satisfies the equations 
$$
  i[P^{\mu},\varphi(x)]_{\star}=\frac{\partial\varphi(x)}
            {\partial x_{\mu}} ~,
\eqno{(4.59)}$$ 
which is the Heisenberg relations in the form of Moyal star-products. 
From Eq. (4.17), we know that Eq. (4.59) (i.e., Eq. (4.18)) is 
equivalent to Eq. (4.16) for 
the case $\theta^{0i}=0$ of the spacetime noncommutativity. This means 
that Heisenberg relations (4.16) in the form of the commutation 
relations of ordinary products are satisfied for noncommutative 
$\varphi^{\star4}$ scalar field theory. Thus we have proved that 
Poincar{\' e} translation invariance is satisfied for noncommutative 
$\varphi^{\star4}$ scalar field theory from the Heisenberg and quantum 
motion equation approach. However we have seen that for our proof, we 
need the condition $\theta^{0i}=0$ of the spacetime noncommutativity, 
i.e., we need the spacetime noncommutativity to be spacelike. For the 
four-momentum $P^{\mu}$ to be conserved quantities of motion for 
noncommutative $\varphi^{\star4}$ scalar field theory, the equivalence 
between Eq. (4.59) and (4.16), and the satisfying of Eq. (4.52), they 
all need $\theta^{0i}$ to be zero. This means that for noncommutative 
$\varphi^{\star4}$ scalar field theory, its Poincar{\' e} translation 
invariance can only be held for spacelike noncommutativity of the 
spacetime. On the other hand, it was shown in Refs. [19-21] that the 
unitarity of noncommutative field theories will also lose when 
$\theta^{0i}\neq 0$. For the light-like noncommutativity, i.e., for the 
case $\theta^{0i}=-\theta^{1i}$ with all other components of 
$\theta^{\mu\nu}$ to be zero, it was argued in Ref. [22] that the 
unitarity of noncommutative field theories are still held. However 
according to the study of this paper, the Poincar{\' e} translation 
invariance for noncommutative field theories may not hold for 
light-like noncommutativity of the spacetime either.

  Because the demonstration of the Poincar{\' e} translation invariance 
for noncommutative $\varphi^{\star4}$ scalar field theory in this paper 
depends on the definition of the energy-momentum tensor of Eq. (3.12), 
while we have seen in Eq. (3.13) that the energy-momentum tensor of 
Eq. (3.12) is not locally conserved, this makes the demonstration for the 
Poincar{\' e} translation invariance of noncommutative $\varphi^{\star4}$ 
scalar field theory of this paper not very ideal. A remedy for this 
problem is to exert a constraint equation for the energy-momentum tensor: 
$$
  \frac{\lambda}{4!}\left[\left[\varphi,
            \partial_{\mu}\varphi\right]_{\star},
            \varphi^{\star 2}\right]_{\star}=0 ~.
  \eqno{(4.60)}  $$  
This will make the energy-momentum tensor of Eq. (3.12) to be locally 
conserved necessarily. However this will bring more complications for 
the classical and quantum motions for the noncommutative 
$\varphi^{\star4}$ scalar field theory.

\section{Discussion}  

\indent

  In the above sections, we have studied the Moyal commutators and 
Poincar{\' e} translation invariance for noncommutative scalar field 
theory. In noncommutative spacetime we can regard the Moyal star-product 
as the basic product operation. Thus we need to study the commutators of 
Moyal star-products for quantum fields on noncommutative spacetime.  
In Sec. II of this paper, we analyzed the Moyal commutators for 
scalar fields. We find that because of the noncommutativity of spacetime 
coordinates, the Moyal commutators are not $c$-number functions. In order 
to obtain the $c$-number results for Moyal commutators, we need to 
evaluate their vacuum state or non-vacuum state expectation values. We 
obtain that the expectation values of Moyal commutators are equal to the 
corresponding commutators in ordinary commutative spacetime. Then from 
the expectation values of the Moyal commutators for scalar fields, we 
analyzed the microcausality problem for noncommutative scalar 
field theory. We find that the expectation values of the Moyal 
commutators are zero if two spacetime points are separated by a spacelike 
interval. Therefore the microcausality is held for the linear operator 
$\varphi(x)$ of free scalar field on noncommutative spacetime. For the 
quadratic operators of free scalar field on noncommutative spacetime 
such as $\varphi(x)\star\varphi(x)$, their microcausality properties 
need to be studied further. Some of their results have been obtained 
in Refs. [10,11]. In addition, the microcausality problem 
discussed here is only restricted to free fields. Because for 
noncommutative field theories, there exist the UV/IR mixing 
problems [12,13]. The infrared singularities that come from non-planar 
diagrams may need people to invoke certain nonlocal terms in 
the renormalization of noncommutative field theories. These nonlocal 
terms may destroy the microcausality for quantum field 
theories on noncommutative spacetime. In Ref. [6], through supposing 
the spectral measure to be the form of $SO(1,1)\times SO(2)$ invariance, 
the authors obtained that microcausality is violated for quantum 
fields on noncommutative spacetime generally. However, such a 
conclusion is a necessary result of the breakdown of the Lorentz 
invariance in Ref. [6].

  Poincar{\' e} translation invariance is a fundamental property for 
quantum field theories. For quantum field theories in ordinary spacetime, 
Poincar{\' e} translation invariance is verified through the 
satisfying of Heisenberg relations (4.16). For noncommutative field 
theories, when the four-momentum $P^{\mu}$ are expanded according to the 
Moyal star-product (2.2), there will exist infinite expansion terms 
hence it is impossible for us to verify the Poincar{\' e} translation 
invariance for quantum field theories on noncommutative spacetime from 
Eq. (4.16). Under the condition that $P^{\mu}$ are conserved quantities 
of motion, or $\theta^{0i}=0$ if $P^{\mu}$ are not conserved quantities 
of motion, we can rewrite the Heisenberg relations in the form of 
Moyal commutators. Thus we obtained Eq. (4.18). From Eq. (4.18) and 
to use the vacuum expectation values for the Moyal commutators, we have 
proved the Poincar{\' e} translation invariance for noncommutative 
$\varphi^{\star4}$ scalar field theory. However we need the spacetime 
noncommutativity to be spacelike in various occasions. This 
means that for noncommutative $\varphi^{\star4}$ scalar field theory, 
its Poincar{\' e} translation invariance can only be held for 
$\theta^{0i}=0$ of the spacetime noncommutativity. On the other hand, 
it was shown in Refs. [19-21] that the unitarity of noncommutative field 
theories will also be lost when $\theta^{0i}\neq 0$. For the light-like 
noncommutativity, it was argued in Ref. [22] that the unitarity of 
noncommutative field theories are still held. However according to the 
demonstration of this paper, Poincar{\' e} translation invariance may 
not hold generally for noncommutative field theories for the light-like 
noncommutativity of the spacetime.

  For the noncommutative parameters $\theta^{\mu\nu}$, usually people 
take them to be constants which are antisymmetric to the indexes 
$\mu$ and $\nu$. We can see that the spacetime noncommutative relations 
(2.1) and (2.7) are translation invariant. The Lagrangian (2.3) is also 
translation invariant. This make us be able to define the energy-momentum 
tensor according to the method of Sec. III. However the energy-momentum 
tensor obtained in Sec. III is not divergence free. This means that for 
noncommutative field theories, the translation invariance of the 
Lagrangians and actions may not necessarily result the existence of 
locally conserved energy-momentum tensors. From the classical field 
equation approach, in Ref. [9] the authors have demonstrated the 
Poincar{\' e} translation invariance for noncommutative field theories. 
In addition to take $\theta^{\mu\nu}$ to be a $c$-number second-order 
antisymmetric tensor, in Ref. [9] the authors have demonstrated the 
Lorentz rotation invariance for noncommutative field theories from their 
classical field equation approach. If we take $\theta^{\mu\nu}$ to be 
a $c$-number second-order antisymmetric tensor in this paper, we can 
same prove the Poincar{\' e} translation invariance for noncommutative 
scalar field theory. This is because $\theta^{\mu\nu}$ is invariant 
under a translation transformation even though it is a second-order 
antisymmetric tensor. The spacetime noncommutative relations (2.1) and 
(2.7) and the Lagrangian (2.3) are still translation invariant if 
$\theta^{\mu\nu}$ is a second-order antisymmetric tensor. Therefore 
the demonstration for the Poincar{\' e} translation invariance of  
noncommutative scalar field theory in this paper is still held for 
$\theta^{\mu\nu}$ to be a second-order antisymmetric tensor.

  We have also generalized the Heisenberg evolution equation (4.1) to 
the form of Eq. (4.9) of Moyal star-products. However because the 
Hamiltonian defined as in Eqs. (4.5) and (4.6) for noncommutative field 
theories are not conserved quantities of motion generally, we need the 
condition $\theta^{0i}=0$ for the spacetime noncommutativity generally 
to obtain Eq. (4.9) for noncommutative field theories. Under the 
condition $\theta^{0i}=0$ for the spacetime noncommutativity, we can 
regard Eq. (4.9) as the fundamental evolution equation for quantum fields 
on noncommutative spacetime. We can establish the $S$-matrix from 
Eq. (4.9) in the interaction picture, so that to keep the field operators 
in the form of Moyal star-products in ${\cal H}_{int}$. Then from the 
corresponding Wick's theorem of field operators of Moyal star-products, 
we can also explain why there are non-planar diagrams in noncommutative 
field theories [4,23]. Many problems for the quantum equation of motion 
of noncommutative field theories have also been studied in Ref. [24] 
from the different approaches and methods. For the Poincar{\' e} 
translation invariance of other noncommutative field theories, we can 
also expect to construct the proofs using the method of this paper. It 
can also be generalized to the demonstration of the Lorentz rotation 
invariance for noncommutative field theories. However, we can expect 
that they will be rather complicated. A different method for the  
understanding of the relativistic invariance of noncommutative field 
theories is the twisted Poincar{\' e} algebra approach [7,8]. However, 
the quantum field realization of such an approach is not very clear at 
present.

\newpage

\centerline{{\large {\bf APPENDIX: A DERIVATION FOR EQUATION (4.52)}}}

\vskip 0.2cm

\indent

  In this Appendix we give a derivation for Eq. (4.52):
$$
  \int d^{3}x\delta^{3}({\bf x}-{\bf x}^{\prime})\star f({\bf x},t)=
  \int d^{3}x f({\bf x},t)\star\delta^{3}({\bf x}-{\bf x}^{\prime})=
   f({\bf x}^{\prime},t) ~. 
  \eqno{({\rm A}1)}$$
The Fourier integral representation of 
$\delta^{3}({\bf x}-{\bf x}^{\prime})$ is given by 
$$
  \delta^{3}({\bf x}-{\bf x}^{\prime})=\frac{1}{(2\pi)^{3}}\int
    e^{i{\bf k}\cdot({\bf x}-{\bf x}^{\prime})}d^{3}k ~.        
  \eqno{({\rm A}2)}$$ 
Under the condition $\theta^{0i}=0$ of the spacetime noncommutativity, 
the Moyal star-products and the integral operations in Eq. ({\rm A}1) do 
not relate with the time variable, so we can write the Fourier integral 
for $f({\bf x},t)$ as
$$
  f({\bf x},t)=\int d^{3}q f({\bf q},t)e^{i{\bf q}\cdot{\bf x}} ~.    
  \eqno{({\rm A}3)}$$ 
To substitute Eqs. ({\rm A}2) and ({\rm A}3) into Eq. ({\rm A}1), we 
obtain 
\begin{eqnarray*}
  & ~ &  \int d^{3}x\delta^{3}({\bf x}-{\bf x}^{\prime})\star 
      f({\bf x},t)        \\
  & = & \frac{1}{(2\pi)^{3}}\int d^{3}x\int
    e^{i{\bf k}\cdot({\bf x}-{\bf x}^{\prime})}d^{3}k\star
    \int d^{3}q f({\bf q},t)e^{i{\bf q}\cdot{\bf x}}       \\
  & = & \frac{1}{(2\pi)^{3}}\int d^{3}q f({\bf q},t) \int d^{3}k
        \int d^{3}x e^{i{\bf k}\cdot({\bf x}-{\bf x}^{\prime})}
        \star e^{i{\bf q}\cdot{\bf x}}         \\
  & = & \frac{1}{(2\pi)^{3}}\int d^{3}q f({\bf q},t) \int d^{3}k 
        \int d^{3}x e^{-\frac{i}{2}k\times q}e^{i{\bf k}\cdot
         ({\bf x}-{\bf x}^{\prime})+i{\bf q}\cdot{\bf x}}       \\
  & = & \frac{1}{(2\pi)^{3}}\int d^{3}q f({\bf q},t) \int d^{3}k
        e^{-\frac{i}{2}k\times q}e^{-i{\bf k}\cdot{\bf x}^{\prime}}
        \int d^{3}x e^{i({\bf k}+{\bf q})\cdot{\bf x}}     \\
  & = & \int d^{3}q f({\bf q},t)\int d^{3}k\delta^{3}({\bf k}+{\bf q})
        e^{-\frac{i}{2}k\times q}
        e^{-i{\bf k}\cdot{\bf x}^{\prime}}                \\
  & = & \int d^{3}q f({\bf q},t)e^{i{\bf q}\cdot{\bf x}^{\prime}}
        e^{\frac{i}{2}q\times q}=f({\bf x}^{\prime},t) ~,
\end{eqnarray*}
$$  \eqno{({\rm A}4)}$$ 
where in the above derivation $k\times q=k_{i}\theta^{ij}q_{j}$. 
Similarly we have 
$$
  \int d^{3}x f({\bf x},t)\star\delta^{3}({\bf x}-{\bf x}^{\prime})=
   f({\bf x}^{\prime},t) ~. 
  \eqno{({\rm A}5)}$$
Thus we have proved Eq. (4.52) to be satisfied under the condition 
$\theta^{0i}=0$ of the spacetime noncommutativity. In the case 
$\theta^{0i}\neq0$, in the third line of Eq. ({\rm A}4), we cannot 
move $f({\bf q},t)$ into the front directly because it will take 
part in the Moyal star-product operation, hence we cannot obtain the 
last result or Eq. ({\rm A}1).

\newpage

\noindent {\large {\bf References}}

\vskip 12pt

\noindent 

[1] S. Doplicher, K. Fredenhagen, and J.E. Roberts, Phys. Lett. B 
    {\bf 331}, 39 (1994); 

    ~~~ Commun. Math. Phys. {\bf 172}, 187 (1995), hep-th/0303037. 

[2] N. Seiberg and E. Witten, J. High Energy Phys. 09 (1999) 032,
hep-th/9908142, 

~~~ and references therein.

[3] M.R. Douglas and N.A. Nekrasov, Rev. Mod. Phys. {\bf 73}, 977 
    (2001), hep-th/0106048.

[4] R.J. Szabo, Phys. Rep. {\bf 378}, 207 (2003), hep-th/0109162.  

[5] H.S. Snyder, Phys. Rev. {\bf 71}, 38 (1947). 

[6] L. {\' A}lvarez-Gaum{\' e} and M.A. V{\' a}zquez-Mozo, Nucl. Phys.
{\bf B668}, 293 (2003), 

~~~ hep-th/0305093. 

[7] M. Chaichian, P. Kulish, K. Nishijima, and A. Tureanu,
    Phys. Lett. B {\bf 604}, 98, 

    ~~~ (2004) hep-th/0408069. 

[8] M. Chaichian, P. Pre{\v s}najder, and A. Tureanu, Phys. Rev. Lett. 
{\bf 94}, 151602 (2005), 

~~~ hep-th/0409096. 

[9] R. Banerjee, B. Chakraborty, and K. Kumar, Phys. Rev. D {\bf 70},
125004 (2004), 

~~~ hep-th/0408197.  

[10] O.W. Greenberg, hep-th/0508057. 

[11] M. Chaichian, K. Nishijima, and A. Tureanu, Phys. Lett. B 
     {\bf 568}, 146 (2003), 

~~~~~ hep-th/0209008. 

[12] S. Minwalla, M. Van Raamsdonk, and N. Seiberg, J. High Energy Phys. 
     02 (2000) 

~~~~~  020, hep-th/9912072. 

[13] M. Van Raamsdonk, J. High Energy Phys. 11 (2001) 006, hep-th/0110093.  

[14] A. Gerhold, J. Grimstrup, H. Grosse, L. Popp, M. Schweda, and 
     R. Wulkenhaar, 

     ~~~~~ hep-th/0012112.

[15] A. Micu and M.M. Sheikh-Jabbari, J. High Energy Phys. 01 (2001) 025, 

~~~~~ hep-th/0008057.  

[16] A. Das and J. Frenkel, Phys. Rev. D {\bf 67}, 067701 (2003), 
     hep-th/0212122.  

[17] L. M{\" o}ller, J. High Energy Phys. 10 (2004) 063, hep-th/0409085.  

[18] J.D. Bjorken and S.D. Drell, {\sl Relativistic Quantum Fields}  
     (McGraw-Hill, 1965). 

[19] J. Gomis and T. Mehen, Nucl. Phys. {\bf B591}, 265 (2000), 
hep-th/0005129.

[20] A. Bassetto, F. Vian, L. Griguolo, and G. Nardelli, 
J. High Energy Phys. 07 (2001) 

~~~~~ 008, hep-th/0105257. 

[21] L. Alvarez-Gaum{\' e}, J.L.F. Barb{\' o}n, and R. Zwicky, 
     J. High Energy Phys. 05 (2001) 

~~~~~ 057, hep-th/0103069.  

[22] O. Aharony, J. Gomis, and T. Mehen, J. High Energy Phys. 09 
(2000) 023, 

~~~~~ hep-th/0006236. 

[23] T. Filk, Phys. Lett. B {\bf 376}, 53 (1996). 

[24] P. Heslop and K. Sibold, Eur. Phys. J. C {\bf 41}, 545 (2005), 
hep-th/0411161.

\end{document}